# MLET: A Power Efficient Approach for TCAM based, IP Lookup Engines in Internet Routers

Hamidreza Mahini[1], Reza Berangi[2] and Alireza Mahini[3]

[1]Department of ELearning, Iran University of Science and Technology (IUST), Tehran, Iran
h_mahini@vu.iust.ac.ir

[2]Department of Computer Engineering, Iran University of Science and Technology (IUST), Tehran, Iran
rberangi@iust.ac.ir

[3]Department of Computer Engineering, Islamic Azad University-Gorgan branch, Gorgan, Iran
mahini@comp.iust.ac.ir

## Abstract

*Routers are one of the important entities in computer networks specially the Internet. Forwarding IP packets is a valuable and vital function in Internet routers. Routers extract destination IP address from packets and lookup those addresses in their own routing table. This task is called IP lookup. Internet address lookup is a challenging problem due to the increasing routing table sizes. Ternary Content-Addressable Memories (TCAMs) are becoming very popular for designing high-throughput address lookup engines on routers: they are fast, cost-effective and simple to manage. Despite the TCAMs speed, their high power consumption is their major drawback. In this paper, Multilevel Enabling Technique (MLET), a power efficient TCAM based hardware architecture has been proposed. This scheme is employed after an Espresso-II minimization algorithm to achieve lower power consumption. The performance evaluation of the proposed approach shows that it can save considerable amount of routing table's power consumption.*

## 1. Introduction

Forwarding of The Internet Protocol (IP) packets is the primary purpose of Internet routers [1]. The speed at which forwarding decisions are made at each router or "hop" places is a fundamental limit on the performance of the network. For Internet Protocol Version 4 (IPv4), the forwarding decision is based on a 32-bit destination address carried out in each packet's header. The use of Classless InterDomain Routing (CIDR) complicates the lookup process, requiring a lookup engine to search a route table containing variable-length address prefixes in order to find the longest matching prefix for the destination address in each packet header and retrieve the corresponding forwarding information. In high-performance routers, each port employs a separate LPM search engine.

As physical links speed grow and the number of ports in high-performance routers continue to increase, there is a growing need for efficient lookup algorithms and effective implementations of those algorithms. Next generation routers must be able to support thousands of optical links 32 each operating at 10 Gb/s (OC-192) or more [2]. Lookup techniques that can scale efficiently to high speeds and large lookup table sizes are essential for meeting the growing performance demands, while maintaining acceptable per-port costs. Many techniques are available to perform IP address lookups. Perhaps the most common approach in high-performance systems is to use Ternary Content Addressable Memory (TCAM) devices. While this approach can provide excellent performance, the performance comes at a fairly high price due to the exorbitant power consumption and high cost per bit of TCAM relative to commodity memory devices [3].Today's high-density TCAMs consume 12 to 15 W per chip when the entire memory is enabled [2]. To support the super linearly increasing number of IP prefixes in core routers, vendors use up to eight TCAM chips. Filtering and packet classification would also require additional chips. The high power consumption of using many chips increases cooling costs and also limits the router design to fewer ports [4]. Recently, researchers have proposed a few approaches to reduce power consumption in TCAMs [4, 5], including routing-table compaction [6, 7]. Liu presents a novel technique to eliminate redundancies in the routing table [3]. However, this technique takes excessive time for update because it is based on the Espresso-II minimization algorithm [8], which exponentially increases in complexity with the number of prefixes in a routing table. Proposed approach in [2] is a TCAM-based architecture that consumes less power than previous works. Additionally, their approach minimizes the memory size required for storing the prefixes, but there is a black box called PEB (Page Enable Block) in their architecture which is more complex to design and the authors of [2] didn't propose any internal design for it. Thus, our work's main objective is a TCAM-based router architecture that consumes less power and is suitable for the incremental updating that modern IP routers need using the routing-table compaction and partitioning with the others.

## 2. RELATED WORKS

IP lookup techniques are divided into three categories as follow: software, hardware and hybrid approaches. In software methods, a data structure named trie is used. In computer science, a trie, or prefix tree, is an ordered tree data structure that is used to store an associative array where the keys are usually strings. One of the best software methods is DVSBC-PC which was presented by Kai Zheng and his colleagues [9]. DVSBC-PC is abbreviation of Dynamic Variable Stride Bitmap Compression and Path Compression. Although this approach combines two techniques to achieve efficient trie for lookup scheme, but it needs 7.17 memory access for IPv4 lookup in average case. This method provider combined it with TCAM and presented a new approach named DVSBC-PC-CAM. This is one instance of hybrid approaches. With this combination they are achieve to 4.33 memory access for IPv4 lookup in average case. Also other hybrid methods are presented. For example some of them use caching for performance improvement [10].

Hardware approaches typically use dedicated hardware for routing lookup [11, 12]. More popular techniques use commercially available content-addressable memory (CAM). CAM storage architectures have gained popularity because their searching time is $O(1)$ that is; it is bounded by a single memory access. Binary CAMs allow only fixed-length comparisons and are therefore unsuitable for longest-prefix matching. The TCAM solves the longest-prefix problem and is by far the fastest hardware device for routing. In contrast to TCAMs, ASICs that use trie— digital trees for storing strings (in this case, the prefixes) [1]—require four to six memory accesses for a single route lookup and thus have higher latencies. Also, TCAM-based routing table updates have been faster than their trie based counterparts. The number of routing-table entries is increasing super linearly [1]. Today, routing tables have approximately 500,000 entries [2], so the need for optimal storage is also very important. Yet CAM vendors claim to handle a maximum of only 8,000 to 128,000 prefixes, taking allocators and deal locators into account [1].The gap between the projected numbers of routing-table entries and the low capacity of commercial products has given rise to work on optimizing the TCAM storage space by using the properties of lookup tables [6, 7]. Even though TCAMs can store large numbers of prefixes, they consume large amounts of power, which limits their usefulness. Panigrahy and Sharma introduced a paged-TCAM architecture to reduce power consumption in TCAM routers[5].Their scheme partitions prefixes into eight groups of equal size; each resides on a separate TCAM chip. A lookup operation can then select and enable only one of the eight chips to find a match for an incoming IP address. In addition, the approach introduces a paging scheme to enable only a set of pages within a TCAM. However, this approach achieves only marginal power savings at the cost of additional memory and lookup delay. Other work describes two architectures, bit selection and trie-based, which use a paging scheme as the basis for a power-efficient TCAM [4]. The bit selection scheme extracts the 16 most significant bits of the IP address and uses a hash function to enable the lookup of a page in the TCAM chip. The approach assumes the prefix length to be from 16 to 24 bits. Prefixes outside this range receive special handling; the lookup searches for them separately. However, the number of such prefixes in today's routers is very large (more than 65,068 for the bbnplanet router) [2], and so this approach will result in significant power consumption. In addition, the partitioning scheme creates a trie structure for the routing table prefixes and then traverses the trie to create partitions by grouping prefixes having the same sub prefix. The sub prefixes go into an index TCAM, which further indexes into static RAM to identify and enable the page in the data TCAM that stores the prefixes. The index TCAM is quite large for smaller page sizes and is a key factor in power consumption. The three-level architecture, though pipelined, introduces considerable delay. It is important to note that both the approaches [4, 5] store the entire routing table, which is unnecessary overhead in terms of memory and power. Although the existing approaches reduce power either by routing table compaction or selecting a portion of the TCAM, our approach reduces power by combining the two approaches, we reduce routing table with logic minimization algorithm and select the suitable partition of TCAM table using our novel technique, MLET (Multilevel Enabling Technique).

## 3. PROPOSED APPROACH

With using the prefix properties and Espresso-II algorithm we reduce the IP lookup table's rows as described earlier .Here, we propose an architectural technique that reduces the IP lookup table laterally. This technique adopts the multi-level routing lookup architecture applying the Multi Stage TCAMs, that we are called MSTCAM. Multilevel Enabling technique (MLET), a power efficient TCAM based hardware architecture is employed after an Espresso-II minimization algorithm to achieve lower power consumption. The performance evaluation of the proposed approach on Telstra routing table shows that it can save up to considerable percentage of routing table's power consumption.

### 3.1. Routing table minimization

Here, we reduce the table entry using Espresso-II algorithm but since the complexity of Espresso-II is directly related with the entries of the algorithm, we use the prefix overlapping technique as described in [2] for reducing the Espresso-II entries. Although the prefix overlapping and minimization techniques together compact about 30 to 45 percent of the routing table [2], these techniques have an overhead when it comes to prefixes that require fast updates. The time taken for prefix overlapping is bounded and independent of the router's size [2]. For table size reduction we used three following techniques sequentially.
 1- Elimination of overlapped prefixes.
 2- Partitioning the result of previous step into PRTs (Partial Routing Tables).
 3- Minimizing the PRTs using Espresso Minimization Units (EMU) in parallel.
Figure1 depicted the schematic of minimization part of our architecture.
As shown in figure1 input of MU is the routing table, in MU the first step is overlap elimination, then the splitter splits the result of overlap minimization into the PRTs via the partitioning raw as described earlier, after partitioning, PRTs are reduced with EMUs simultaneously. Each EMU is as same as ROCM which is proposed by Vahid and Lysecky in [13]. Finally the Merger unit merges the reduced PRTs and makes the Minimized routing table as the output of MU. For supporting of incremental updates the other duty of merger is saving begin and end addresses of each PRT in minimized routing table. In the next sections we describe the overlap elimination process and table partitioning raw.

### 3.1.1. Overlap elimination

The overlap elimination technique eliminates redundant routing prefixes. We use $|P_a|$ to denote the length of prefix $P_a$, and use $P_{a,i}$ to denote the $i^{th}$ bit of the prefix, where $P_{a,1}$ is the most significant bit and $P_{a,|Pa|}$ is the least significant bit. A prefix $P_a$ is the parent of prefix $P_b$ if the following three conditions hold:
   1. $|P_a| < |P_b|$.
   2. $P_{a,i} = P_{b,i}$ for all $1 < i < |P_a|$.
   3. There is no prefix $P_c$ such that $|P_a| < |P_c| < |P_b|$, and $P_{c,i} = P_{b,i}$ for all $1 \leq i \leq |P_c|$.

Intuitively, the parent of prefix $P_b$ is the longest prefix that matches the first few bits of $P_b$. A parent $P_a$ of prefix $P_b$ is an identical parent if $P_a$ translates to the same route as $P_b$—that is, packets matching both prefixes will be routed to the same next hop. The idea of overlap elimination is fairly simple. If $P_a$ is an identical parent of $P_b$, then $P_b$ is a redundant routing prefix. To understand this, assume the longest prefix matched for an IP address is $P_b$; by definition, the IP address will match $P_a$ as well. With $P_b$ removed from the routing table, $P_a$ becomes the longest matched prefix. Because they both translate to the same route, removing $P_b$ makes no difference. Note that this technique is general enough that it applies to any routing lookup algorithm, regardless of how the routing table is stored. In table1 a part of Telestra routing table is shown and in table2 the result of overlap elimination depicted.

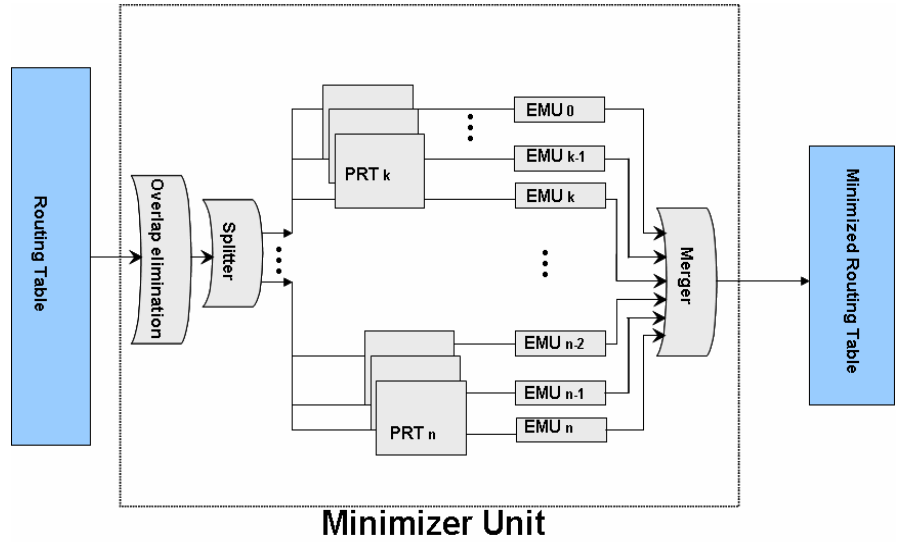

Figure 1.Minimization unit (MU) schema

### 3.1.2. Route Table Partitioning and PRTs

The partitioning rule is very straightforward. It partitions the prefixes based on their corresponding output port. Suppose $P = \{p_1, p_2, ..., p_m\}$ is the set of $M$ prefixes collected by a backbone router. Assuming there are $N$ output ports (e.g. the line cards of router), let $Q = \{1,2,...N\}$ denote the set of these port indices. A set of all $(p_i, q_i)$ pairs, shown as $L$, indicates a many-to-one function that maps $P$ into $Q$. Thus, a lookup table is an organized set of $e_i = (p_i, q_i)$ pairs. We partition set $L$ into $N$ subsets $L^1$ to $L^N$ such that for partition $L^k$ we have:

$$\forall (p_i^k, q_i^k) \in L^k \qquad q_i^k = k \qquad (1)$$

For minimization we suppose that each subset $L^K$ as a separated table and call it Partial Route Table or PRT.

## 4. Proposed Architecture

In this architecture we propose using of Multistage TCAM array instead of the general TCAM array structure. The proposed architecture is shown in figure2. In this architecture prefixes are stored in the MSTCAM after the minimization process. In a general TCAM table, searching in rows needs to enable all cells of each row simultaneously, either content of cells are matched with search entry or not. However in a MSTCAM table, enabling of row cells is done stage by stage. In other word comparing of search entry and content of rows in a MSTCAM is done stage by stage. At the search process the first stage of rows are enabled and comparison are done for all first stages of rows and if stage$_1$ is matched the next stage of row will be enabled. In other word any $M$ bits row table can be divided to $K$ stages which each stage has $W_i$ word length if and only if the following relation is satisfied:

$$\sum_{i=1}^{k} W_i = M, (1 \le k \le M) \qquad (2)$$

International Journal of Computer Networks & Communications (IJCNC), Vol.2, No.3, May 2010

Table 1. Telestra route table entries

| Prefix | Next hop Id |
|---|---|
| 0100000001101001 | 0 |
| 0100000001101010 | 0 |
| 0100000001101011 | 0 |
| 0100001010100111 | 1 |
| 0100001010101000 | 1 |
| 0100001010101001 | 1 |
| 0100001010101011 | 1 |
| 0100000100001010 | 2 |
| 0100000100001011 | 2 |
| 0100000100001011 | 2 |
| 0100000100001110 | 2 |
| 0100000001101101 | 0 |
| 0100000001101110 | 0 |
| 0100000001101111 | 0 |
| 0100000001101111 | 0 |
| 0100001010101100 | 1 |
| 0100000100001111 | 2 |
| 0100000100010010 | 2 |
| 0100000100010011 | 2 |
| 0100001010101101 | 1 |
| 0100001010101110 | 1 |
| 0100000001110000 | 0 |
| 0100000001110001 | 0 |
| 0100000100010110 | 2 |
| 0100000100010111 | 2 |
| 0100000100100000 | 2 |
| 0100000100101000 | 2 |
| 0100000001110010 | 0 |
| 0100000001110011 | 0 |
| 0100000001110100 | 0 |
| 0100000001110101 | 0 |
| 0100000100101111 | 2 |
| 0100000100111100 | 2 |
| 0100000001110110 | 0 |
| 0100000001111010 | 0 |

Table 2. Result of overlap elimination on table1

| Prefix | Next hop Id |
|---|---|
| 01000000011-1010 | 0 |
| 0100000001101--1 | 0 |
| 0100000001101-1 | 0 |
| 0100000001110-0 | 0 |
| 010000000111 00 | 0 |
| 0100000001110--0 | 0 |
| 0100001010100111 | 1 |
| 010000101010 11-0 | 1 |
| 0100001010101 0-1 | 1 |
| 0100001010101-0 | 1 |
| 0100000100111100 | 2 |
| 010000010010-000 | 2 |
| 0100000100010-1 | 2 |
| 0100000100-01111 | 2 |
| 0100000100001-1 | 2 |

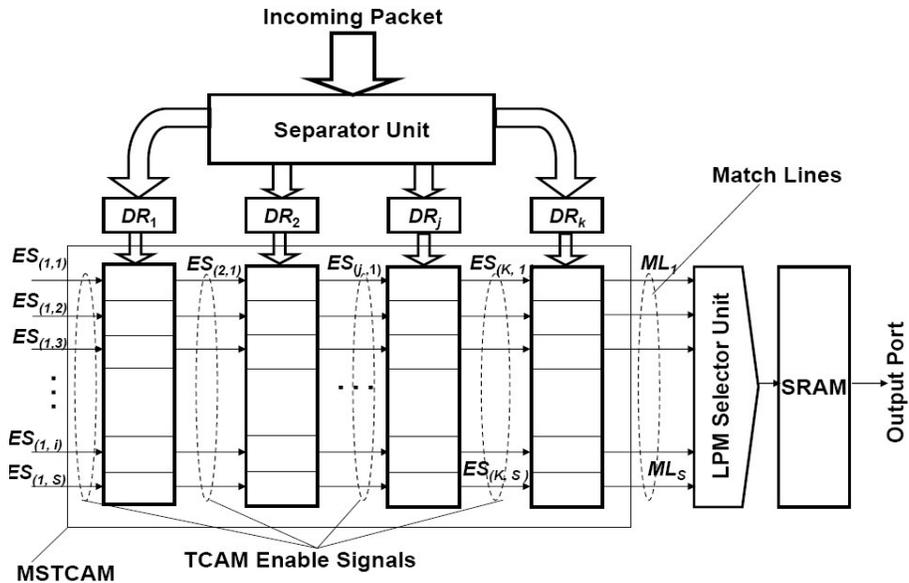

Figure 2. Proposed architecture

The separator unit (SU) extracts the destination IP address of incoming packet. As we described comparison in each stage must be done with the separated part of IP address, thus the IP address must be split to *K* parts which each part length is equal to the bit length of related stage. This splitting mechanism also is done by separator unit.

After IP address extraction and splitting, the split parts of IP address is stored into the Data Registers (DRs).

ES signals are Enabling Signals of each stage in rows. In other word enabling the stage$_i$ of row$_j$ is depend on the $ES_{(i,j)}$ activation.

ML signals are Match Line signals which are placed in the output of the last stage determine which row is matched with destination IP address.

The LPM selector unit, dose the Longest Prefix Matching selection.

### 4.1. Lookup operation

When a new packet is received at one of the input port of router the SU extract the destination IP address from the IP header and prepare it for each stage by split the address and store split parts in DRs. Lookup begins with activation of $ES_{(1,i)}$, thus the first stage is enabled and $W_1$ bits of most significant bits of IP address which are stored in $DR_1$, are compared with all rows in stage$_1$ simultaneously. If the content of row$_i$ in first stage is matched with $DR_1$ then the $ES_{(2,i)}$ is activated and it means that the comparison will continue in stage$_2$ but if the content of row$_i$ in first stage is not matched with $DR_1$ comparison stopped and other bits of this row wont enable. Therefore with this method we don't enable unnecessary TCAM cells and it is equal to power consumption.

If the lookup table has *S* rows and *K* stages we have:

1- $\forall i, j, i = 1, 1 \le j \le S, : ES_{(i,j)} = 1$ .

2- $\forall i, j, 1 < i < k, 1 \le j \le S, : ES_{(i+1,j)} = Match_{(i,j)}$ .

3- $\forall i, j, i = k, 1 \le j \le S, : ML_j = Match_{(i,j)}$

4- The time complexity of lookup operation is *O(k)*.

Figure3 shows the lookup operation by an activity diagram.

### 4.2. Update Operation

Approximately 100 to 1,000 updates per second take place in core routers today [14].Thus, the update operation should be incremental and fast to avoid becoming a bottleneck in the search operation. Update operation include two sub operations: insert and withdrawal. Our main objective in update operation is that the minimum part of TCAM table to be influenced. In other approaches for example in [6] all rows of the TCAM table or a large size of it will be involved with the prefixes compaction after update operation. Minimizing such a large set of prefixes introduces two types of delays: The first delay comes from Espresso-II's computation time. The second delay comes from TCAM entry updates. For any routing update, our approach restricts TCAM updates to a related PRT. So it is possible to update several PRTs simultaneously using multiple EMUs. Our technique can achieve a higher number of updates per second than what a single TCAM chip supports by using several TCAM chips and placing each on a separate bus.

### 4.3 Insert

Suppose that a new prefix must be added to the table. It means that the pair $e_i = (p_i, q_i)$ must be added to $L^{q_i}$ set. So, according to the output port of $p_i$ prefix which is $q_i$, the start and end addresses of $PRT_{q_i}$ got from Merger Unit (MU) then the PRT entries with new $p_i$ sent to the related EMU and the result information stored in MU. Finally rebuilding of the table and saving the new addresses must be done. Figure 4 depicted the activity diagram of insert operation.

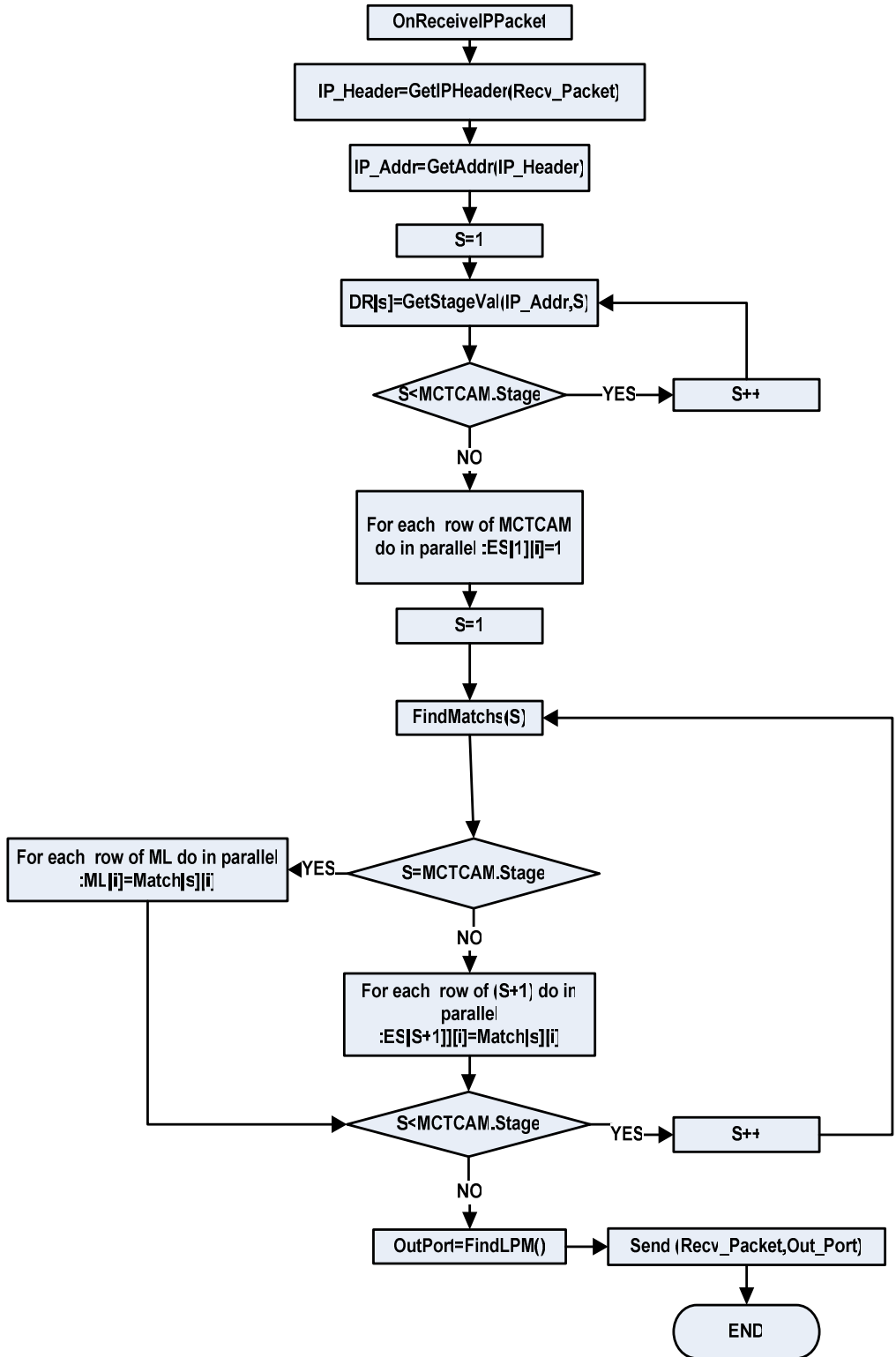

Figure 3. Lookup activity diagram

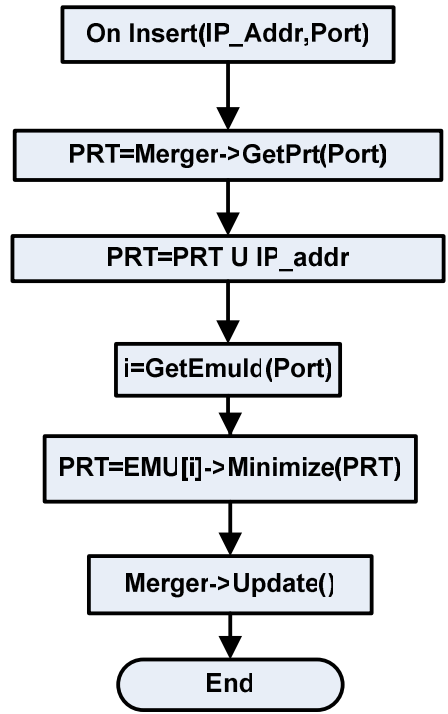

Figure 4. Insert activity diagram

## 4.4 Withdrawal

The algorithm for removing a prefix from the routing table is more complex because several cubes could cover the prefix. Because of the PRT after passing from EMU includes the cover cubes we must remove all cubes covering the prefix and recalculate a minimum cover from the affected prefixes. Figure 5 provides an example. $C_1$, $C_2$, and $C_3$ are cubes, and $P_1$, $P_2$, $P_3$, and $P_4$ are prefixes. If $P_3$ needs to be removed, then $C_2$ and $C_3$ must be removed. As a result, $P_2$ and $P_4$ no longer have any cover, so they must be included in the computation for new cover. Note that $P_1$ isn't affected, because although $C_2$ is removed, $C_1$ still covers $P_1$. The incremental removal algorithm search for prefixes no longer covered by cubes and includes them in the computation for new cover. The activity diagram for the incremental removal algorithm appears in Figure 6.

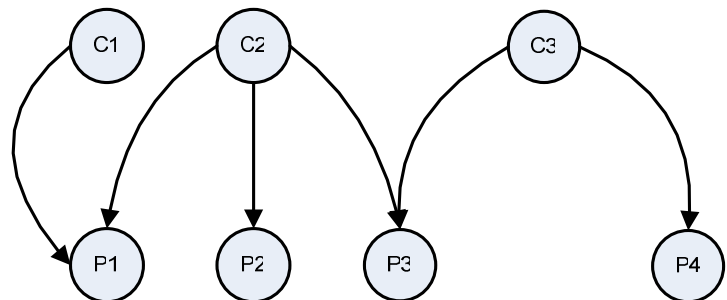

Figure 5. Cube and prefix relationship

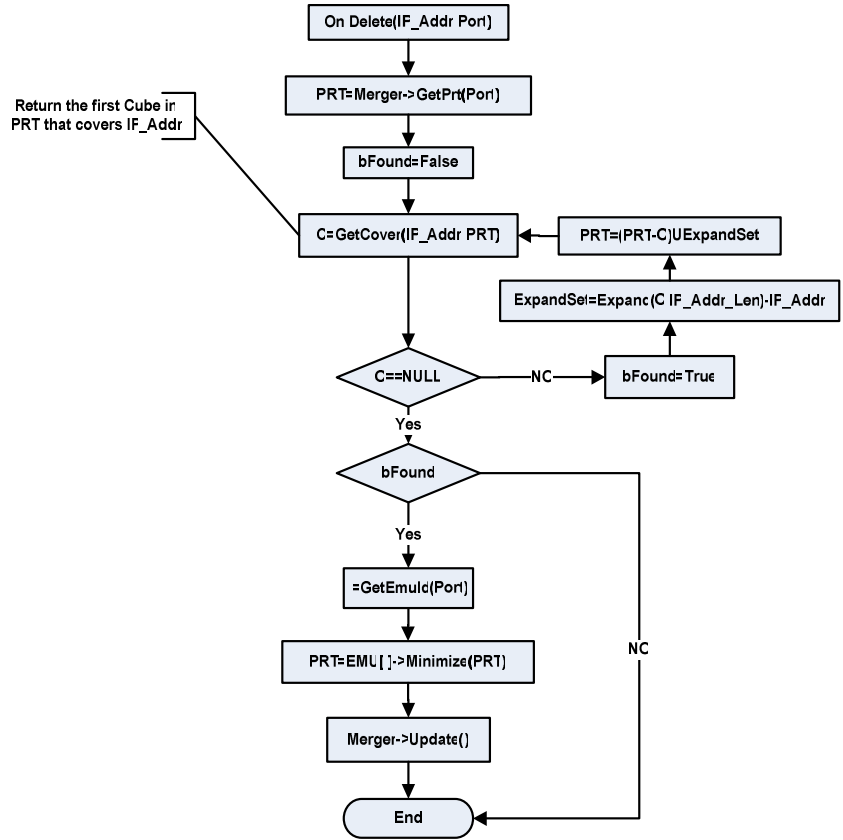

Figure 6. Withdrawal activity diagram

## 5. PERFORMANCE EVALUATION

In this section we want to evaluate our approach performance. Thus we define some performance metrics and terms as below.

- *EPS (Enabled bits Per Search)*: power consumption in lookup operation directly depends on the number of enabled bits of TCAM table cells, so EPS can be a good parameter for measuring the power consumption. According to the applied architecture, EPS can be constant or variable in each search. For example if the reference model [15] is used, we have:

$EPS = S \times W$  (3)

Where $S$ is the number of TCAM rows and $W$ is the bit length of each row. Lookup operation in reference model needs to enable all cells of TCAM table and this is worst case for power consumption so, EPS of this model is the maximum and we call it $EPS_{max}$.

- *MEPS (Mean Enabled bits Per Search):* This term refers to the mean of EPSs for a set of search so we can say that: if lookup is done for m addresses MEPS obtain from following formula:

$MEPS = \sum_{i=1}^{m} EPS_i / m$  (4)

Note that the maximum MEPS belongs to the reference model and is equal to $EPS_{max}$ so we have:

$MEPS_{max} = EPS_{max}$  (5)

- *POF (Power Optimization Factor)*: This parameter refers to the power optimization percentage which is defined as the relation of mean power optimization in each search to the maximum power consumption per each search. Suppose that a TCAM cell consume $P$ watt of power when it is enabled so the POF of search for $m$ address in TCAM table obtain from formula 4:

21

$$POF = \frac{(MEPS_{max} - MEPS) \times p}{MEPS_{max} \times p} \times 100 = (1 - \frac{\sum_{i=1}^{m} EPS_i}{EPS_{max}}) \times 100 = (1 - \frac{\sum_{i=1}^{m} EPS_i}{S \times W}) \times 100 \quad (4)$$

Where *S* refers to the row number of table and *W* refers to the bit length of each row.
Therefore for reference model we have: POF=0. Now we can evaluate current approach with calculating POF of them.

In our approach MSTCAM is used and according to the lookup operation described in section 4.1 surely our MEPS is less than MEPS$_{max}$. Note that in our approach the enabled bits in TCAM table depend on both of the stage number in MSTCAM and the bit length of each stage. For example in the figure 7 the advantage of using MSTCAM is visible. In this figure the enabled cells are bold. As you see in the figure 7 if the stages increase the enabled cells decrease and its effect of MLET technique.

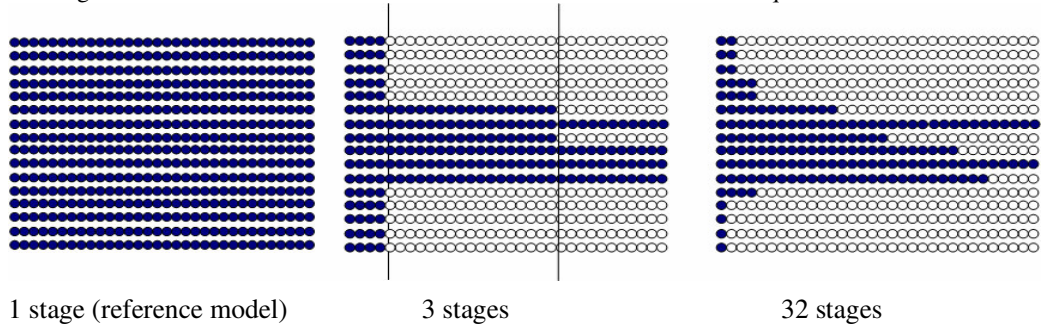

1 stage (reference model)       3 stages              32 stages
Figure 7. Relation of stage number and enabled cells

For more careful evaluation we simulate our approach and use it for Telestra routing table.

### 5.1. Experimental Results

We will describe the results in two aspects: the first is results of our minimization technique and the second is results of using MLET.

#### 5.1.1. Minimization results

In our simulation 31000 prefixes of the existing prefixes in Telestra are given to the minimization unit as input set and after minimization process the prefix set decrease to the 12372 prefixes. Thus we could compact the TCAM table about 60 percentage. Note that the minimization result of Ravikumar and Mahaparta in [2] is similar to ours because our approach and theirs are very similar in minimization rule.

#### 5.1.2 MLET Results

For making clear the advantage of using this technique, we examined the following configurations for the minimized table:
   a. 2 stages patterns ($w_1, w_2$).
   b. 3 stages patterns ($w_1, w_2, w_3$).
   c. 4 stages patterns ($w_1, w_2, w_3, w_4$).
   d. K stages patterns where K is the power of 2 and $w_1 = w_2 = ... = w_k = \frac{W}{k}, 1 \leq k \leq 32$.

The following results obtain from testing the 10000 address from incoming address list of Telestra router which are selected randomly. In the figure 8 you can see the POF diagram of all cases of 2 stages configuration. The horizontal bar represents the word length of the first stage and the POF value of this word length maps to the vertical bar.

International Journal of Computer Networks & Communications (IJCNC), Vol.2, No.3, May 2010

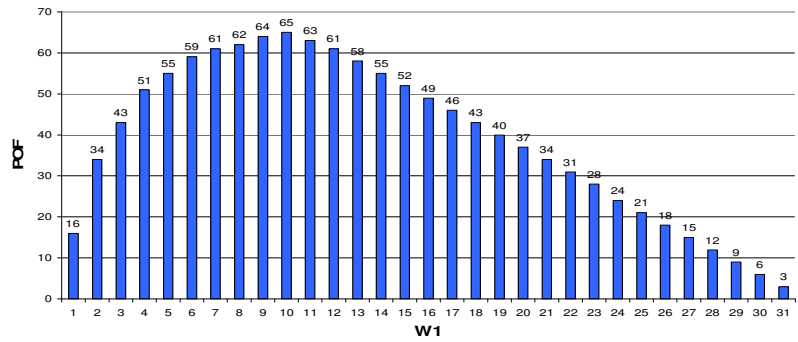

Figure 8. POF diagram of 2 stages configuration

In the figure 9, 10 you can see the POF diagram of all cases of 3 and 4 stages configuration. The horizontal bar represents the word length of the first stage and the maximum POF value of this word length maps to the vertical bar.

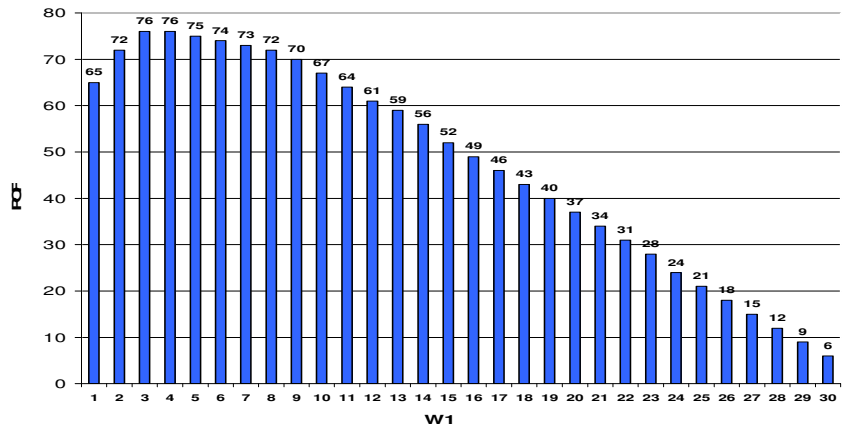

Figure 9. POF diagram of 3 stages configuration

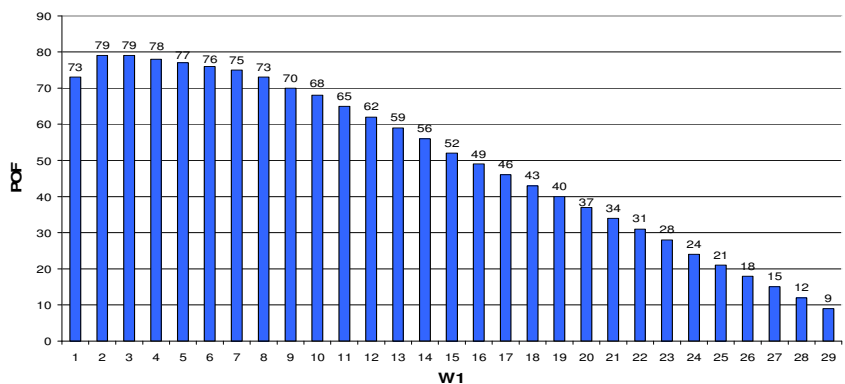

Figure 10. POF diagram of 4 stages configuration

The last model is the power of 2 stages model where the word length of all stages in each configuration are equal. This model is noticeable because of in this model each stage in a configuration is as same as the other. These patterns consist of:
   a. 1 stage where its word length is 32.
   b. 2 stages where its word length is 16.

  c. 4 stages where its word length is 8.
  d. 8 stages where its word length is 4.
  e. 16 stages where its word length is 2.
  f. 32 stages where its word length is 1.

In the figure 11 you can see the POF diagram of all cases of power 2 stages configuration. The horizontal bar represents the number of stages and the POF value of the stage number maps to the vertical bar.

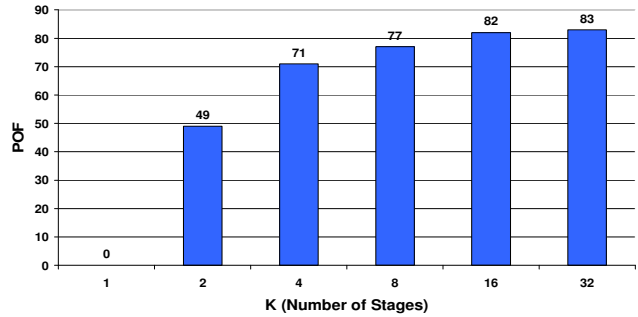

Figure 11.POF diagram of power 2 stages configuration

For comparison of described models, look at the figure 12, please. In figure 12 all of configurations are depicted in a two dimensional diagram which the horizontal bar represents the number of stages and maximum POF value of the stage number maps to the vertical bar.

As you can see in the figure 12 the maximum POF is belong to 32 stages configuration which is 83. But it is true that the increase in the stages lead to increase in implementation complexity. So it seems that 4 stages is more suitable than 32 stage because the number of stage in it is very less than 32 stage in spite of its POF is very little difference with the 32 stages. Anyway there is a tradeoff between power consumption and hardware complexity. To show the performance of our proposed architecture we compare it with available strongly recommended techniques. So we simulated the PEB based [2], LIU [6], Bit Selection [4] and IFPLUT [7] models which are presented by details in[16], for Telestra routing table and this simulation result is depicted in the table3.

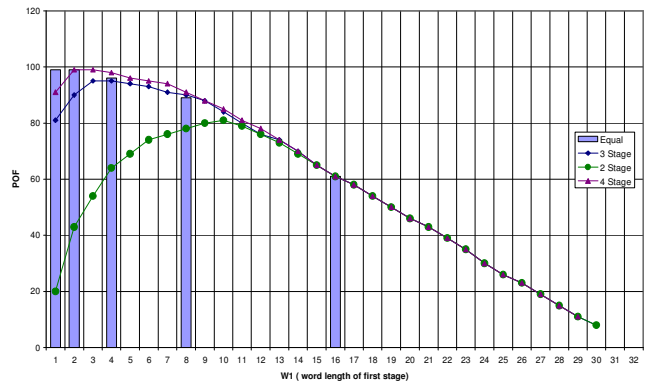

Figure 12.Comparison of different models

Table 3.Performance comparison of available models with ours

|  | MLET | | | | | | | |
|---|---|---|---|---|---|---|---|---|
|  | Equal | 4-Stages | 3-Stages | 2-Stages | PEB | LIU | IFPLUT | Bit Selection |
| Minimization POF | 60 | 60 | 60 | 60 | 58 | 53 | 0 | 0 |
| Enabling POF | 83 | 79 | 76 | 65 | 75 | 0 | 36 | 28 |
| Total POF | 93.2 | 91.6 | 90.4 | 86 | 89.5 | 53 | 36 | 28 |

## 6. CONCLUSION

In this paper, we presented a novel architecture for a TCAM-based IP forwarding engine. We have shown significant reduction in memory usage based on the prefix compaction and architectural design. We designed a heuristic to match entries in TCAM stages so that only a bounded number of entries are looked up during the search operation. A fast incremental update scheme has been introduced that is time bounded. The memory requirements and power consumption for router architecture have been outlined. To demonstrate the merit of the proposed architecture, we used the architectural features on Telestra router based on its trace statistics and evaluated the benefits of our approach. It has been shown that the memory references are reasonably decreased due to the use of Multilevel Enabling Technique (MLET) and effective compaction technique. At the same time, the power consumption is found to be remarkably low to promise efficient TCAM design in the future.

**Authors**

**Hamidreza Mahini,** received the B.S. degree in software engineering in 2006 from Iran University of Science and Technology (IUST), Iran. And he received the master degree in ICT engineering in the same university. His current researches focus on high performance IP route lookup, network processor architecture, grid computing, semistructured data models such as XML and Artificial Neural Network.

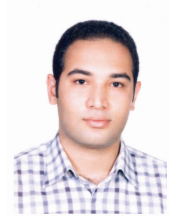

**Reza Berangi** received the PhD in Mobile Telecommunications. Now he is an Assistant of Professor at Iran University of Science and Technology (IUST) and head of the wireless network lab in the school of computer engineering in the same university. His current researches focus on high performance IP route lookup, network processor architecture and mobility management for the next generation networks (NGN).

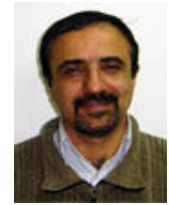

**Alireza Mahini,** received the master degree in computer systems architecture in 2006 from Iran University of Science and Technology (IUST), Iran. Now he is a PhD candidate in the Islamic Azad University, sciences and researches branch. He is academic staff of Islamic Azad University, Gorgan branch and Dean of Gorgan SAMA vocational training center . His current researchs focus on high performance IP route lookup, network processor architecture, and NOC.

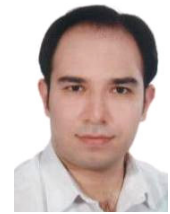